\documentstyle[12pt]{article}
\def\ie{{\em i.e.}}
\def\eg{{\em e.g.}}

\def\beq{\begin{equation}}
\def\eeq{\end{equation}}

\catcode`\@=11 % This allows us to modify PLAIN macros.
\def\coeff#1#2{{\textstyle{#1\over #2}}}

\def\lsim{\mathrel{\mathpalette\@versim<}}
\def\gsim{\mathrel{\mathpalette\@versim>}}
\def\@versim#1#2{\vcenter{\offinterlineskip
    \ialign{$\m@th#1\hfil##\hfil$\crcr#2\crcr\sim\crcr } }}

\def\JL{J. L. Lopez}
\def\DVN{D. V. Nanopoulos}

\def\r#1{$\bf#1$}
\def\rb#1{$\bf\overline{#1}$}

\def\t1{{\tilde 1}}

\def\mpt{p\hskip-5.5pt/\hskip2pt}

\def\GeV{\,{\rm GeV}}
\def\TeV{\,{\rm TeV}}
\def\y{\,{\rm y}}

\def\to{\rightarrow}
\def\pb{\,{\rm pb}}
\def\ipb{\,{\rm pb}^{-1}}
\def\NPB#1#2#3{Nucl. Phys. B {\bf#1} (19#2) #3}
\def\PLB#1#2#3{Phys. Lett. B {\bf#1} (19#2) #3}

\def\PRD#1#2#3{Phys. Rev. D {\bf#1} (19#2) #3}
\def\PRL#1#2#3{Phys. Rev. Lett. {\bf#1} (19#2) #3}
\def\PRT#1#2#3{Phys. Rep. {\bf#1} (19#2) #3}

\def\TAMU#1{Texas A \& M University preprint CTP-TAMU-#1}

\begin{document}
% TH format
\begin{flushright}
\baselineskip=12pt
{CERN-TH.7077/93}\\
{CERN/LAA/93-29}\\
{CTP-TAMU-65/93}\\
{ACT-22/93}\\
\end{flushright}
% PPE format
%\begin{center}
%{\large EUROPEAN ORGANIZATION FOR NUCLEAR RESEARCH}
%\end{center}
%\begin{flushright}
%{CERN-PPE/93-??}\\
%{?? November, 1993}\\
%{CERN-LAA/93-29}\\
%{CERN-TH.7077/93}\\
%{CTP-TAMU-65/93}\\
%{ACT-22/93}\\
%\end{flushright}
%

\begin{center}
\vglue 0.3cm
{\Large\bf A Layman's Guide to SUSY GUTs\\}
\vspace{0.2cm}
\vglue 0.5cm
{JORGE L. LOPEZ$^{(a),(b)}$, D. V. NANOPOULOS$^{(a),(b),(c)}$, and A.
ZICHICHI$^{(d)}$\\}
\vglue 0.4cm
{\em $^{(a)}$Center for Theoretical Physics, Department of Physics, Texas A\&M
University\\}
{\em College Station, TX 77843--4242, USA\\}
{\em $^{(b)}$Astroparticle Physics Group, Houston Advanced Research Center
(HARC)\\}
{\em Mitchell Campus, The Woodlands, TX 77381, USA\\}
{\em $^{(c)}$CERN Theory Division, 1211 Geneva 23, Switzerland\\}
{\em $^{(d)}$CERN, 1211 Geneva 23, Switzerland\\}
\baselineskip=12pt

\vglue 0.5cm
{\tenrm ABSTRACT}
\end{center}
%\vglue 1cm
{\rightskip=3pc
 \leftskip=3pc
%\xpt\baselineskip=12pt
\noindent
The determination of the most straightforward evidence for the existence
of the Superworld requires a guide for non-experts (especially
experimental physicists) for them to make their own judgement on the value of
such predictions. For this purpose we review the most basic results of
Super-Grand unification in a simple and clear way. We focus the attention on
two specific models and their predictions. These two models represent an
example of a direct comparison between a traditional unified-theory and
a string-inspired approach to the solution of the many open problems of the
Standard Model. We emphasize that viable models must satisfy {\em all}
available experimental constraints and be as simple as theoretically possible.
The two well defined supergravity models, $SU(5)$ and  $SU(5)\times U(1)$, can
be described in terms of only a few parameters (five and three respectively)
instead of the more than twenty needed in the MSSM model, \ie, the Minimal
Supersymmetric extension of the Standard Model. A case of special interest is
the strict no-scale $SU(5)\times U(1)$ supergravity where all predictions
depend on only one parameter (plus the top-quark mass). A general consequence
of these analyses is that supersymmetric particles can be at the verge of
discovery, lurking around the corner at present and near future facilities.
This review should help anyone distinguish between well motivated
predictions and predictions based on arbitrary choices of parameters in
undefined models.
}
% TH format
\vspace{1cm}
\begin{flushleft}
\baselineskip=12pt
{CERN-TH.7077/93}\\
{CTP-TAMU-65/93}\\
{ACT-22/93}\\
October 1993
\end{flushleft}
\vfill\eject
\tableofcontents
\vfill\eject
\setcounter{page}{1}
\pagestyle{plain}
\baselineskip=14pt

\section{Introduction}
\label{Introduction}
The purpose of this review paper is to present a simple guide for
non-specialists into the complex world of Supersymmetry.  A world with many
appealing features: the boson-fermion equivalence, the unification of all gauge
couplings and masses, consequently of all forces, including gravity, and the
ultimate goal of discovering the Theory of Everything (TOE). The large number
of papers published on this subject does not allow the possibility of everyone
carefully distinguishing unjustified claims from real achievements, especially
for those cases of experimental interest. It is precisely in this field that
clarity is badly needed. To predict the energy level where the Superworld
should show up is one of the most exciting problems of modern physics. The
effort needed to implement a project, in terms of people and  financial
resources, is so vast that experimentalists themselves need to judge the value
of a ``prediction".  For example, if a paper is published where the energy
level is shown to be unaccessible with present facilities, those physicists
engaged for years in related experiments should be able to understand and judge
the real value of such predictions. The primary purpose of the present paper is
to achieve this goal.

\section{Two supergravity models: why these choices?}
\label{why}
The LEP $e^+e^-$ collider at CERN has been in operation since 1989 with a
center-of-mass energy around the $Z$-boson mass ($M_Z\approx91\GeV$).
During this time a vast amount of data has been collected on many different
decay modes of the $Z$. Analyses of the totality of the data in the context of
the Standard Model of electroweak interactions show no deviations from
expectations, which have been tested at the level of one-loop and for some
quantities two-loop corrections \cite{Altarelli}. In particular, the weak
mixing angle ($\sin^2\theta_W(M_Z)$) and the QCD running coupling
($\alpha_3(M_Z)$) have been determined with unprecedented accuracy.  This
picture-perfect agreement between theory and experiment has had several
consequences. One of these is the immediate rejection of large classes of
models which predicted new physics below the electroweak scale.

The lesson from LEP is therefore how to reconcile the successes of the Standard
Model with the compelling reasons for the existence of new physics beyond it.
An obvious shortcoming of the Standard Model is the ad-hoc nature of the many
parameters involved: the
quark and lepton masses, the quark mixing angles, and the CP violating phase.
However, even if explanations for these parameters could be found within
physics not so far from the electroweak scale, a problem would show up when
extrapolating the theory to much larger energies. This so-called {\em gauge
hierarchy problem} is manifest in theories with elementary scalar fields, such
as the Higgs boson in the Standard Model. The reason is that radiative
corrections to the Higgs-boson mass become extremely large if the theory also
contains very massive particles, as is the case in unified theories, or any
theory which attempts to incorporate gravity in a common framework. This is a
``problem" for the theory, as opposed to a terminal disease, in that there is a
grossly undesirable solution which involves fine-tuning parameters to large
numbers of decimal places, order-by-order in perturbation theory.

Particle physics seems to always find solutions to its problems by invoking
the existence of larger symmetries. The proposed solutions to the above two
problems of the Standard Model are no different: unified models and
supersymmetry, respectively. The former postulate that the Standard Model
is actually embedded in a larger group structure which includes $SU(3)_C\times
SU(2)_L\times U(1)_Y$ but which is only fully manifest at very large energies,
where new degrees of freedom are excited and the symmetry group is effectively
enlarged to \eg, $SU(5),SU(5)\times U(1),SO(10)$. In the larger theory
relations among the parameters arise, which help explain some of the
regularities observed in the Standard Model parameters, such as the
quantization of the electric charge and some relations among the quark and
lepton masses. In fact, the existence of three light generations of quarks and
leptons had already been predicted in this context in studies of the
$m_b/m_\tau$ ratio \cite{BEGN+NR}. An immediate consequence of the larger
symmetry is that the gauge couplings of the Standard Model are ``unified", \ie,
take the same value at the unification scale ($E_{GUT}$), even though they are
measured to be different numbers at the weak scale. By-products of great
experimental interest are the new interactions among the quarks and leptons
which may be induced by the heavy degrees of freedom. Among these, the most
model-independent one leads to the prediction of proton decay, which must occur
at a rather slow rate ($\tau_p\gsim10^{32}\y$) to avoid conflict with
experimental limits.

Supersymmetry solves the gauge hierarchy problem by predicting the existence
of partners for the ordinary particles which differ by half-a-unit of spin,
\eg, electron($1\over2$)-selectron($0$), photon($1$)-photino($1\over2$).
The radiative corrections to the Higgs-boson mass are now tamed by the fact
that the partners of the heavy particles have an opposite effect on the
correction. However, since supersymmetry cannot be an unbroken symmetry
(otherwise the sparticles would have the same mass as their Standard Model
partners and no such sparticles have been observed) the cancellation is not
perfect, but up to the typical mass splitting of particles and their
superpartners. If this splitting can be predicted to be no larger than
$\sim1\TeV$, then the gauge hierarchy problem will remain solved. In fact, it
is this further requirement which demands that we enlarge the symmetry even
more, by making supersymmetry a local symmetry called {\em supergravity}. In
this class of theories the mass splittings can be explicitly computed in terms
of very few parameters. This ability is of great experimental interest, since a
large class of (\eg, collider and rare) low-energy processes can be calculated
in terms of few parameters, providing close knit correlations among the several
experimental predictions, which may otherwise appear arbitrary.

We will consider two such supergravity models: the minimal $SU(5)$
\cite{Dickreview} and $SU(5)\times U(1)$ \cite{EriceDec92}. Conceptually,
the $SU(5)$ model is a typical (and the simplest) grand unified supersymmetric
model which predicts the existence of proton decay at a rate which should be
on the verge of being observed. In fact, its non-supersymmetric ancestor, the
Georgi-Glashow $SU(5)$ model \cite{GG}, was ruled out experimentally on the
basis of its incorrect (too short) prediction for the proton lifetime.
Theoretically, this model also suffers from the gauge hierarchy problem, which
is present in the absence of supersymmetry. In the supersymmetric $SU(5)$
model, the unification of gauge couplings occurs at a scale
$E_{GUT}\sim10^{16}\GeV$, and the proton lifetime problem of non-supersymmetric
$SU(5)$ is naturally solved. As we now well know, unification of the gauge
couplings also does not work in non-supersymmetric $SU(5)$, while it works
very well in its supersymmetric counterpart. The goodness of the prediction for
the $m_b/m_\tau$ relation alluded to above is also maintained with the addition
of supersymmetry \cite{BEGN+NR}.

 The $SU(5)\times U(1)$ supergravity model is best motivated in the context of
superstrings, where it is found that $SU(5)\times U(1)$ (in contrast with
$SU(5)$) is relatively easy to obtain. In this model the unification of
couplings should occur at the string scale ($E_{GUT}\sim10^{18}\GeV$). This
fact appears to require a non-minimal spectrum of particles at intermediate
scales, in contrast with the $SU(5)$ model where the ``big desert" picture is
assumed to hold. Furthermore, the $SU(5)\times U(1)$ supergravity model allows
a further reduction of the unknown parameters of the theory, making its
predictions very sharp. Moreover, these two models allow to compare possible
implementations of supersymmetry: one ($SU(5)$) as example of traditional
unified theories, the other ($SU(5)\times U(1)$) as example of string-inspired
theories.

\section{Convergence of gauge couplings: geometry versus physics}
\label{convergence}
The three gauge couplings of the Standard Model, as measured at the $Z$-boson
mass scale, can be expressed in terms of: $\alpha_e$, $\sin^2\theta_W$, and
$\alpha_3$. Experimentally we know that (see \eg, \cite{LP}),
\begin{eqnarray}
\alpha^{-1}_e(M_Z)&=&127.9\pm0.1,\\
\alpha_3(M_Z)&=&0.120\pm0.010,\\
\sin^2\theta_W(M_Z)&=&0.2324\pm0.0006.
\end{eqnarray}
The $U(1)_Y$ and $SU(2)_L$ gauge couplings are related to these by
$\alpha_1={5\over3}(\alpha_e/\cos^2\theta_W)$ and
$\alpha_2=(\alpha_e/\sin^2\theta_W)$. In numbers:
\begin{equation}
{\rm At}\quad Q=M_Z\qquad\left\{
\begin{array}{ccc}
\alpha^{-1}_1&=&58.91\pm0.16\\
\alpha^{-1}_2&=&29.72\pm0.080\\
\alpha^{-1}_3&=&8.33\pm0.69
\end{array}
\right.\label{initialconditions}
\end{equation}
The three gauge couplings evolve with increasing values of the scale $Q$ in
a logarithmic fashion, and may become equal at some higher scale, signaling
the possible presence of a larger gauge group. However, this need not be the
case: the three gauge couplings may meet and then depart again. Conceptually,
the presence of a unified group is essential in the discussion of unification
of couplings. In this case, the newly excited degrees of freedom will be
such that all three couplings will evolve together for scales $Q>E_{GUT}$, and
one can then speak of a unified coupling. (For a recent review and extensive
references to the current literature see \eg, Ref.~\cite{Erice93}.)

The running of the gauge couplings is prescribed by a set of first-order
non-linear differential equations: the renormalization group equations (RGEs)
for the gauge couplings. In general, there is one such equation for each
dynamical variable in the theory (\ie, for each gauge coupling, Yukawa
coupling, and sparticle mass). These equations give the rate of change of each
dynamical variable as the scale $Q$ is varied. For the case of a gauge
coupling, the rate of change is proportional to (some power of) the gauge
coupling itself, and the coefficient of proportionality is called the {\em beta
function}. The beta functions encode the spectrum of the theory, and how the
various gauge couplings influence the running of each other (a higher-order
effect). Assuming that all supersymmetric particles have a common mass
$M_{SUSY}$, the RGEs (to two-loop order) are:
\begin{equation}
{d\alpha^{-1}_i\over
dt}=-{b_i\over2\pi}-\sum^3_{j=1}{b_{ij}\alpha_j\over8\pi^2}.\label{dgidt}
\end{equation}
where $t=\ln (Q/E_{GUT})$, with $Q$ the running scale and $E_{GUT}$ the
unification mass. The one-loop ($b_i$) and two-loop ($b_{ij}$) beta functions
are given by
\begin{eqnarray}
b_i&=&\left(\coeff{33}{5},1,-3\right), \\
b_{ij}&=&\left( \begin{array}{c@{\quad}c@{\quad}c}
{199\over 25} & {27\over 5} & {88\over 5} \\
{9\over 5} & 25 & 24 \\ {11\over 5} & 9 & 14
\end{array} \right).
\end{eqnarray}
These equations are valid from $Q=M_{SUSY}$ up to $Q=E_{GUT}$. For
$M_Z<Q<M_{SUSY}$
an analogous set of equations holds, but with beta functions which reflect
the non-supersymmetric nature of the theory (\ie, with all the sparticles
decoupled),
\begin{eqnarray}
b'_i&=&\left(\coeff{41}{10},-\coeff{19}{6},-7\right), \\
b'_{ij}&=&\left( \begin{array}{c@{\quad}c@{\quad}c}
{199\over 50} & {27\over 10} & {44\over 5} \\
{9\over 10} & {35\over6} & 12 \\ {11\over 10} & {9\over2} & -26
\end{array} \right).
\end{eqnarray}
The non-supersymmetric equations are supplemented with the initial conditions
given in Eq.~(\ref{initialconditions}).

If the above is all the physics which is incorporated in the study of the
convergence of the gauge couplings, then it is easy to see that the couplings
will always meet at some scale $E_{GUT}$, provided that $M_{SUSY}$ is tuned
appropriately \cite{EKN,LL,AdBF,ACPZ}. This is a simple consequence of
euclidean geometry, as can be seen from Eq.~(\ref{dgidt}). Neglecting the
higher-order terms, we see that as a function of $t$, $\alpha^{-1}_i$ are just
straight lines. In fact, the slope of these lines changes at $Q=M_{SUSY}$,
where the beta functions change. The convergence of three straight lines with a
change in slope is then guaranteed by euclidean geometry, as long as the point
where the slope changes is tuned appropriately. (This fact was pointed out by
A. Peterman
and one of us (A. Z.) in 1979 \cite{X1}.) What is non-trivial about the
convergence of the couplings is that with the initial conditions given in
Eq.~(\ref{initialconditions}), the change in slope needs to be
$M_{SUSY}\sim1\TeV$ \cite{AdBF}.

\begin{figure}[p]
\vspace{6.5in}
\includegraphics{fig1Layman.ps}
\vspace{0.5cm}
\caption{\baselineskip=12pt
The convergence of the gauge couplings $(\alpha_1, \alpha_2, \alpha_3)$ at
$E_{GUT}$ is followed by the unification in a unique $\alpha_{GUT}$ above
$E_{GUT}$. The RGEs include the heavy and light thresholds plus the evolution
of gaugino masses. These results are obtained using as input the world-average
value of $\alpha_3(M_Z)$ and comparing the predictions for $\sin^2\theta(M_Z)$
and $\alpha_{em}^{-1}(M_Z)$ with the experimental results.  The $\chi^2$
constructed using these two physically measured quantities allows to get the
best $E_{GUT}$, $\alpha_{GUT}(E_{GUT})$, and $\alpha_3(M_Z)$ (shown). The RGEs
can go down to $M_Z$ without any need of introducing a change of slope at
$E\approx10^3$ GeV as
%was the case for the ``geometrical" fit of Ref. [10].
would be required if the various effects mentioned above are neglected.}
\label{FigureY2b}
\end{figure}

This prediction \cite{AdBF} for the likely scale of the supersymmetric spectrum
(\ie, $M_{SUSY}\sim1\TeV$ \cite{AdBF}) is in fact {\em incorrect} \cite{EGM}.
The reason is simple:  the physics at the unification scale, which is used to
predict the value of $M_{SUSY}$, has been ignored completely. In fact, such a
geometrical picture of convergence of the gauge couplings is physically
inconsistent, since for scales $Q>E_{GUT}$ the gauge couplings will depart
again. One must consider a unified theory to be assured that the couplings will
remain unified, as shown in Fig.~\ref{FigureY2b}. This entails the study of a
new kind of effect, namely the influence on the running of the gauge couplings
of the degrees of freedom which are excited near the unification scale (\ie,
the {\em heavy threshold effects}) \cite{EKN,ACZ}. In fact, the whole concept
of a single unification point needs to be abandoned. The upshot of all this is
that the theoretical uncertainties on the values of the parameters describing
the heavy GUT

 particles are such that the above prediction for $M_{SUSY}$ \cite{AdBF} is
washed out completely \cite{EGM}. Furthermore, the insertion of a realistic
spectrum of sparticles at low energies (as opposed to an unrealistic common
$M_{SUSY}$ mass) blurs the issue some more \cite{EKN,EGM}. {\em Thus, it is
perfectly possible to obtain acceptable unification, with  supersymmetric
particle masses as low as experimentally allowed.} The most complete analysis
of a unified theory is shown in Fig.~\ref{FigureY2}. Note the unification of
the gauge couplings which continues above $E_{GUT}$. Notice also that ``light"
and ``heavy" thresholds have been duly accounted for, plus other detailed
effects like the evolution of gaugino masses (EGM). This effect has in fact
been calculated at two loops \cite{X2}.

A related point is that LEP data do not uniquely demonstrate that the gauge
couplings must unify at a scale $E_{GUT}\sim10^{16}\GeV$ \cite{ACZ}. This is
probably the simplest conclusion one could draw. However, this conclusion is
easily altered by for example considering models with particles at intermediate
scales, \ie, by populating the ``big desert". In fact, one such simple
modification allows the gauge couplings to converge at the string scale
$E_{GUT}\sim10^{18}\GeV$ instead \cite{EriceDec92}.

\begin{figure}[p]
\vspace{6in}
\includegraphics{fig2Layman.eps}
\vspace{1cm}
\caption{\baselineskip=12pt
This is the best proof that the convergence of the gauge
couplings can be obtained with $M_{SUSY}$ at an energy level as low as $M_Z$.
Notice that the effects of ``light" and ``heavy" thresholds have been
accounted for, as well as the Evolution of Gaugino Masses [14,15].  This figure
is Fig. 2 of ref. [16]. $E_{SU}$ is the string unification scale.}
\label{FigureY2}
\end{figure}

\section{Constraints from unification}
\label{constraints}
The convergence of the gauge couplings implies that given $\alpha_e$ and
$\alpha_3$, one is able to compute the values of $\sin^2\theta_W$, the
unification scale $E_{GUT}$, and the unified coupling $\alpha_U$. In
lowest-order
approximation (\ie, neglecting all GUT thresholds, two-loop effects, and
taking $M_{SUSY}=M_Z$) one obtains
\begin{eqnarray}
\ln{E_{GUT}\over
M_Z}&=&{\pi\over10}\left({1\over\alpha_e}-{8\over3\alpha_3}\right),\label{MU}\\
{\alpha_e\over\alpha_U}&=&{3\over20}\left(1+{4\alpha_e\over\alpha_3}\right),
\label{au}\\
\sin^2\theta_W&=&0.2+{7\alpha_e\over15\alpha_3}\quad.\label{sin2}
\end{eqnarray}
These equations provide a rough approximation to the actual values obtained
when all effects are included. Nonetheless, they embody the most important
dependences on the input parameters. In Fig.~\ref{Figure1} (from
Ref.~\cite{ACPZ}) we show the relation between $E_{GUT}$ and $\alpha_3$ for
various values of $\sin^2\theta_W$. One can observe that:
\begin{eqnarray}
\alpha_3\ \uparrow\quad &\Rightarrow&\quad E_{GUT}\ \uparrow
\qquad{\rm for\ fixed}\ \sin^2\theta_W\label{a3zMu}\\
\sin^2\theta_W\ \uparrow\quad &\Rightarrow&\quad E_{GUT}\ \uparrow
\qquad{\rm for\ fixed}\ \alpha_3\\
\alpha_3\ \uparrow\quad &\Rightarrow&\quad \sin^2\theta_W\ \downarrow
\qquad{\rm for\ fixed}\ E_{GUT}
\end{eqnarray}
These are the most important systematic correlations, which are not really
affected by the neglected effects. These correlations are evident in
Eqs.~(\ref{MU}--\ref{sin2}) and in Fig.~\ref{Figure1}. In this figure
we also show the lower bound on $E_{GUT}$ which follows from the proton
decay constraint. Clearly a lower bound on $\alpha_3(M_Z)$ results, which
allows the world-average value. Another interesting result is the
anticorrelation between $E_{GUT}$ and $M_{SUSY}$. This is shown in
Fig.~\ref{Figure2}, where for fixed $\sin^2\theta_W(M_Z)$ we see that
increasing $\alpha_3(M_Z)$ increases $E_{GUT}$ (as already noted in
Eq.~(\ref{a3zMu})) and decreases $M_{SUSY}$. Taking for granted this approach
(\ie, all supersymmetric particle masses degenerate at $M_{SUSY}$)
for comparison with the large amount of papers published following this
logic, in Fig.~\ref{Figure3} we see the narrow band left open once the
experimental limits on $\tau_p$ and $M_{SUSY}$ are imposed.
Figure~\ref{Figure3} is a guide to understand the qualitative interconnection
between the basic experimentally measured quantities, $\alpha_3(M_Z)$,
$\sin^2\theta_W(M_Z)$, $\tau_p$, $M_{SUSY}$ and the theoretically desirable
$E_{GUT}$. The experimental lower bounds on the proton lifetime
$(\tau_p)_{exp}$ and on $M_{SUSY}$ produce two opposite bounds (lower and
upper, respectively) on the unification energy
scale $E_{GUT}$.  Note that, in order to make definite predictions
on the lightest detectable supersymmetric particle, a detailed supergravity
model is needed. The study of the correlations between the basic quantities, as
exemplified in this figure, is interesting but should not be mistaken as
example of prediction for the superworld. In particular, the introduction of
the quantity $M_{SUSY}$ is really misleading.

\begin{figure}[p]
\vspace{7.7in}
\includegraphics{fig3Layman.ps}
\vspace{0.5cm}
\caption{\baselineskip=12pt
The unification scale $E_{GUT}$ versus $\alpha_3(M_Z)$ for various values of
$\sin^2\theta_W(M_Z)$ within $\pm2\sigma$ of the world-average value. Also
indicated is the lower bound on $E_{GUT}$ from the lower limit on the proton
lifetime.}
\label{Figure1}
\end{figure}

\begin{figure}[p]
\vspace{7.7in}
\includegraphics{fig4Layman.ps}
%\vspace{0.5cm}
\caption{\baselineskip=12pt
The unification scale $E_{GUT}$ versus $M_{SUSY}$ for different values of
$\alpha_3(M_Z)$ and fixed $\sin^2\theta_W(M_Z)$. Note the anticorrelation
between $M_{SUSY}$ and $E_{GUT}$. The experimental lower bound on $M_{SUSY}$
is shown. The lower bound on $E_{GUT}$ from Fig. 1 is also indicated.}
\label{Figure2}
\end{figure}

\begin{figure}[p]
\vspace{7in}
\includegraphics{fig5Layman.ps}
\vspace{0.5cm}
\caption{\baselineskip=12pt
The correlation between all measured quantities, $\alpha_3(M_Z)$,
$\sin^2\theta_W(M_Z)$, $\tau_p$, the limits on the lightest detectable
supersymmetric particle (here represented by $M_{SUSY}$) and the unification
energy scale $E_{GUT}$.}
\label{Figure3}
\end{figure}

%\section{Where does $M_{SUSY}$ come from?}
\section{The origin of $M_{SUSY}$ and the need for local supersymmetry}
\label{origin}
The calculations which we have described so far, attempted to determine
the value of $M_{SUSY}$, or more properly the supersymmetric particle spectrum,
by fitting the spectrum to obtain the ``best possible" unification picture.
This program did not succeed because of the large inherent uncertainties in
the physics at the GUT scale. Nevertheless, for a given GUT model, it should
be possible to compute the GUT threshold effects and attempt the ``best fit"
procedure to deduce the corresponding light supersymmetric spectrum.
This picture is not very satisfying since one would like to know why the
supersymmetric spectrum should be the way the fit would require it to be. In
other words, the real question is: what determines the values of the sparticle
masses? And why should these be below $\sim1\TeV$, so that the gauge hierarchy
problem is not re-introduced?

As mentioned in Sec.~\ref{why}, considering a theory with supergravity (instead
of global supersymmetry) provides the means to compute the masses of the
sparticles \cite{Dickreview}. This framework
assumes that supersymmetry breaking occurs in a ``hidden sector" of the theory,
where ``gravitational particles" (those introduced when the supersymmetry
was made local) may grow vacuum expectation values (vevs) which break
supersymmetry spontaneously in the hidden sector. These vevs are best
understood as induced dynamically by the condensation of the supersymmetric
partners of the hidden sector particles when the gauge group which describes
them becomes strongly interacting at some large scale. The splitting of the
particles and their partners would then be generated, and would be the order of
the condensation scale ($\sim10^{12-16}\GeV$). However, such huge mass
splittings will not be immediately transmitted to the ``observable" (the
normal) sector of the theory, since the two sectors only communicate through
gravitational interactions. The dampening in the transmission mechanism is such
that the splittings in the observable sector are usually much more suppressed
than those in the hidden sector, and suitable choices of hidden sectors may
yield realistic low-energy supersymmetric spectra. This picture of hidden and
observable sectors becomes completely natural in the context of superstrings,
where models typically contain both sectors and one can study explicitly the
predicted spectrum of supersymmetric particles at low energies.

In a large number of models, the supersymmetric particle masses at the
unification scale are also ``unified". This situation is called {\em universal
soft-supersymmetry-breaking}, and the masses of all scalar partners (\eg,
squarks and sleptons) take the common value of $m_0$, the gaugino (the partners
of the gauge bosons) masses are given by $m_{1/2}$, and there is a third
parameter ($A$) which basically parametrizes the mixing of stop-squark mass
eigenstates at low energies. The breaking of the electroweak symmetry is
obtained dynamically in the context of these models, through the so-called
{\em radiative electroweak symmetry breaking mechanism}, which involves the
top-quark mass in a fundamental way \cite{LN}. After all these well motivated
theoretical ingredients have been incorporated, the models depend on only five
parameters: $m_{1/2}, m_0, A$, the top-quark mass ($m_t$), and the ratio of the
two Higgs vacuum expectations values ($\tan\beta$) \cite{aspects}.

%Having said all this, we still have not answered the question posed by the
%title of this section: where does $M_{SUSY}$ come from?
A rather interesting framework  occurs in the so-called {\em no-scale}
scenario \cite{Lahanas,EKNI+II,LN}, where all the scales in the theory are
obtained from just one basic scale (\ie, the unification scale or the Planck
scale) through radiative corrections. These models have the unparalleled virtue
of a vanishing cosmological constant at the tree-level {\em even after
supersymmetry breaking}, and in their unified versions predict that the
universal scalar masses and trilinear couplings vanish (\ie,
$m_0=A=0$) and the universal gaugino mass $(m_{1/2}$) is the only seed of
supersymmetry breaking. Moreover, this unique mass can be determined in
principle by minimizing the vacuum energy at the electroweak scale. The
generic result is $m_{1/2}\sim M_Z$ \cite{Lahanas,EKNI+II}, in agreement with
theoretical prejudices (\ie, ``naturalness"). Furthermore, no-scale
supergravity is obtained in the infrared limit of superstrings \cite{Witten}.

More generally, in generic supergravity models the five-dimensional parameter
space is constrained by phenomenological requirements, such as sparticle and
Higgs-boson masses not in conflict with present experimental lower bounds,
a sufficiently long proton lifetime, a sufficiently old Universe (a
cosmological constraint on the amount of dark matter in the Universe today),
various
indirect constraints from well measured rare processes, etc. In addition,
further theoretical constraints can be imposed which give $m_0$ and $A$ as
functions of $m_{1/2}$, and thus reduce the dimension of the parameter space
down to just three. In what follows we will focus on some specific supergravity
models which are so constrained that precise experimental predictions can be
made.

\section{Details on the chosen supergravity models: SU(5) and SU(5)xU(1)}
\label{models}
The two supergravity models that we have chosen belong to the class of
models we just described. However, these models are even more predictive
than the run-of-the-mill supergravity model because further well motivated
theoretical assumptions are made.
\subsection{The minimal SU(5) supergravity model}
\label{minsu5}
In this model the gauge group is $SU(5)$, which completely contains the
Standard Model as a subgroup. This implies that all three Standard Model
gauge couplings should evolve from low energies and become equal at (and
above) the unification scale $E_{GUT}$ (up to heavy threshold corrections).
This scale has some dependence on the uncertainties on the various parameters;
one usually obtains $E_{GUT}\sim10^{16}\GeV$. The Standard Model particles and
their superpartners are assigned to the  \rb{5} and \r{10} representations:
\begin{equation}
\bar{\bf5}=\{d^c,L=(\begin{array}{c}e\\ \nu_e\end{array})\},
\qquad {\bf10}=\{Q=(\begin{array}{c}u\\ d\end{array}),u^c,e^c\}.
\end{equation}
The heavy GUT particles which are
excited near the unification scale include the \r{24} representation of
gauge bosons (and gauginos) which contains the twelve Standard Model gauge
bosons (8 gluons, $W^\pm,Z,\gamma$) plus twelve heavy (charged and colored)
gauge bosons which mediate the $SU(5)$ gauge interactions. There is also a
\r{24} representation of Higgs bosons (and their superpartners), and the
vacuum expectation value of the neutral component of this set effects the
$SU(5)\to SU(3)_C\times SU(2)_L\times U(1)_Y$ gauge symmetry breaking.
The minimal $SU(5)$ spectrum also includes a pair of pentaplet representations
(\r{5},\rb{5}) which contain the two Higgs doublets of the low-energy
supersymmetric theory, but also a pair of colored Higgs triplets.

Perhaps the most decisive property of this model is the prediction for the
proton lifetime. Proton decay can be mediated in two ways: by exchange
of heavy gauge or Higgs bosons (dimension-six operators), or by the exchange
of heavy Higgsinos (dimension-five operators) \cite{WSY,ENR}. The dimension-six
operators predict a proton lifetime proportional to $M^4_U$ or $M^4_H$
respectively, and the expected values of these mass scales make this
contribution rather small. On the other hand, the dimension-five operators
entail a proton lifetime proportional to $M^2_{\widetilde H}$, which requires
$M_{\widetilde H}>E_{GUT}$ and some strong constraints on the supersymmetric
spectrum (if it is light enough to be observable). In fact, the
three-dimensional soft-supersymmetry-breaking parameter space ($m_{1/2},m_0,A$)
is constrained in such a way that the squarks and sleptons should be heavy,
while the neutralinos and charginos should be light \cite{ANpd}.

Another important constraint on the parameter space of the minimal $SU(5)$
supergravity model is provided by cosmological considerations. The models
we consider include a discrete symmetry called {\em R-parity} which has
value $+1$ for the ordinary particles, and $-1$ for the supersymmetric
particles. This implies that supersymmetric particles must always be created
or destroyed in pairs, \ie, at each vertex there are always none, two, or four
supersymmetric particles. In particular, the {\em lightest supersymmetric
particle} (LSP) must be stable in these models. Astrophysical considerations
require this particle to be neutral and colorless \cite{EHNOS}. The only two
candidates are the lightest neutralino (a linear combination of the weakly
interacting neutral sparticles: photino, zino, two Higgsinos) and the
sneutrino. It turns out that it is the neutralino which is the lightest one.
Since the neutralinos are stable, they must have pair-annihilated away quite
efficiently in the early Universe, otherwise their present relic abundance
would be too large, and the Hubble parameter would correspond to a Universe
younger than the oldest known stars.

The cosmological constraint can be usually satisfied as long as the
supersymmetric particles are not too heavy. However, the proton decay
constraint requires heavy squarks and sleptons and thus the constraint is
relevant. In fact, the only particles which mediate pair-annihilation
efficiently are the $Z$-boson and the lightest Higgs boson, which are light.
In practice, the neutralino mass must be near half the masses of these
particles so that resonant $s$-channel annihilation occurs \cite{troubles}. The
resulting effect on the parameter space of the model is quite severe
\cite{LNP}.
\subsection{The SU(5)xU(1) supergravity model}
\label{flipped}
The $SU(5)\times U(1)$ (``flipped $SU(5)$") gauge group \cite{Barr} differs
from $SU(5)$ in several ways. The Standard Model particles are also assigned to
the \rb{5} and \r{10} representations, but in a ``flipped" way (\ie,
$u^c\leftrightarrow d^c$, $e^c\leftrightarrow \nu^c$) relative to the minimal
$SU(5)$ case,
\begin{equation}
\bar{\bf5}=\{u^c,L=(\begin{array}{c}e\\ \nu_e\end{array})\},
\qquad {\bf10}=\{Q=(\begin{array}{c}u\\ d\end{array}),d^c,\nu^c\},
\qquad {\bf1}=e^c.
\end{equation}
Note that a right-handed neutrino $\nu^c$ has appeared naturally.
This leads to an automatic {\em see-saw} mechanism to generate small neutrino
masses in this model \cite{revitalized}. The heavy GUT spectrum of fields
includes the \r{24} representation of gauge boson plus the $U(1)$ gauge boson
of $SU(5)\times U(1)$. Unlike $SU(5)$, the symmetry breaking Higgs fields are
contained in the \r{10},\rb{10} representations (which have neutral components
because of the ``flipped" assignment). This property is central to the appeal
of $SU(5)\times U(1)$ as a paradigm of a string model \cite{revamped}, since
larger Higgs representations (like the \r{24}) are not easily obtainable in
string model building. There is also a pair of Higgs pentaplets \r{5},\rb{5}
which contain the two Higgs doublets of the low-energy theory, and the heavy
Higgs triplets.

Proton decay is a concern that is easily dismissed in the typical $SU(5)\times
U(1)$ models: the dimension-six proton decay operators are small as usual,
while the perilous dimension-five operators are strongly suppressed. The latter
is a direct consequence of the $SU(5)\times U(1)$ symmetry as it applies to
the solution of the {\em doublet-triplet splitting problem}. The Higgs triplets
are not only heavy but also cannot mediate the dangerous proton decay
diagram \cite{revitalized,faspects}.

The model we have studied is most simply understood in the context of string
model building when the gauge couplings of the Standard Model unify at the
string scale $E_{GUT}\sim10^{18}\GeV$ \cite{LNZI}. This entails the addition of
another \r{10}, \rb{10} representations to the heavy GUT spectrum, otherwise
the gauge couplings would unify as in the minimal $SU(5)$ model. (Specific
string models with this property have also been constructed \cite{LNY}.)

\begin{table}[b]
\hrule
\caption{The approximate proportionality coefficients to the gluino mass, for
the various sparticle masses in the two supersymmetry breaking scenarios
considered for $SU(5)\times U(1)$ supergravity.}
\label{Table1}
\begin{center}
\begin{tabular}{|c|c|c|}\hline
&no-scale&dilaton\\ \hline
$\tilde e_R,\tilde \mu_R$&$0.18$&$0.33$\\
$\tilde\nu$&$0.18-0.30$&$0.33-0.41$\\
$2\chi^0_1,\chi^0_2,\chi^\pm_1$&$0.28$&$0.28$\\
$\tilde e_L,\tilde \mu_L$&$0.30$&$0.41$\\
$\tilde q$&$0.97$&$1.01$\\
$\tilde g$&$1.00$&$1.00$\\ \hline
\end{tabular}
\end{center}
\hrule
\end{table}

For the soft-supersymmetry-breaking parameters we have used two string-inspired
scenarios, which allow one to determine two of the parameters as functions of
the third one. We have considered,
\begin{description}
\item (i) the {\em no-scale} model:\quad $m_0=A=0$
\item (ii) the {\em dilaton} model:\quad $m_0={1\over\sqrt{3}}m_{1/2}\,,\quad
A=-m_{1/2}$ \cite{KL}.
\end{description}
Note that the dimension of the parameter space is now reduced to just three:
$m_t,\tan\beta$, and $m_{1/2}$, where $m_{1/2}\propto m_{\tilde g}$. This
implies that the sparticle masses are (up to $\tan\beta$-dependent effects)
proportional to the gluino mass. The corresponding approximate proportionality
coefficients are shown in Table~\ref{Table1}. For both these supersymmetry
breaking scenarios there is a yet more restrictive possibility which we call
the {\em strict no-scale} and the {\em special dilaton} scenarios. A further
requirement allows one to deduce the value of $\tan\beta$ as a function of
$m_t$ and $m_{1/2}$, yielding highly predictive two-parameter models.

With these choices for the supersymmetry breaking parameters, one determines
that the cosmological constraint does not restrict the models in any way
\cite{LNZI,LNZII}. In fact, these two models offer a natural explanation
for the required amount of dark matter in the Universe.
%\vfill\eject

\section{Experimental Predictions}
\label{predictions}
As remarked above, in a generic supergravity model one expects a large degree
of correlation among the predictions for, \eg, the sparticle and Higgs
masses and the rates for the various experimental processes of interest.
These generic models are further constrained by the non-observation of any
sparticle or Higgs boson. Interestingly enough, such five-parameter models
are still not very predictive in that many possible trends of correlations
are possible. In contrast, in the two specific supergravity models described in
Sec.~\ref{models}, the correlations become sharp and of much more experimental
interest.

Without imposing any further constraints on the models, besides the present
collider bounds on sparticle and Higgs masses and the proton decay and
cosmological constraints as applicable, we have calculated rates for the
following processes at the indicated facilities:
\begin{eqnarray}
&{\rm Tevatron\ \cite{LNWZ}}&p\bar p\to \chi^\pm_1\chi^0_2,\quad \chi^\pm_1\to
\chi^0_1 l^\pm\nu_l,\quad \chi^0_2\to\chi^0_1 l^+l^-, \quad
l=e,\mu\quad{\rm``trileptons"}\nonumber\\
&{\rm LEP I\ \cite{LNPWZh}}&e^+e^-\to Zh\to f\bar f h\nonumber\\
&{\rm LEP\ II\ \cite{LNPWZ}}&\left\{
\begin{array}{l}
e^+e^-\to \chi^+_1\chi^-_1,\quad \chi^+_1\to \chi^0_1q\bar q',
\quad \chi^-_1\to \chi^0_1 l^-\nu_l,\ l=e,\mu\quad{\rm``mixed\ events"}\\
e^+e^-\to \chi^+_1\chi^-_1,\quad \chi^+_1\to \chi^0_1 l^+\bar\nu_l,
\quad \chi^-_1\to \chi^0_1 l^-\nu_l,\ l=e,\mu\quad{\rm``dilepton\ events"}\\
e^+e^-\to \tilde l^+_R\tilde l^-_R,\quad \tilde l^+_R\to \chi^0_1l^+,
\quad\tilde l^-_R\to \chi^0_1l^-,\quad l=e,\mu,\tau\quad{\rm``dilepton\
events"}\\
e^+e^-\to Z^*h\to f\bar f h,\quad h\to \chi^0_1\chi^0_1,b\bar b,c\bar
c,\tau^+\tau^-,gg
\end{array}\right.\nonumber\\
&{\rm HERA\ \cite{hera}}&\left\{
\begin{array}{l}
e^-p\to\tilde e^-_R\chi^0_1+p,\quad \tilde e^-_R\to \chi^0_1e^-
\quad{\rm``elastic\ selectron-neutralino"}\\
e^-p\to\tilde \nu_e\chi^-_1+p,\quad \chi^-_1\to \chi^0_1 e^-\bar\nu_e
\quad{\rm``elastic\ sneutrino-chargino"}\\
e^-p\to\tilde e^-_R\chi^0_1+X,\quad \tilde e^-_R\to \chi^0_1e^-
\quad{\rm``deep-inelastic\ selectron-neutralino"}\\
e^-p\to\tilde \nu_e\chi^-_1+X,\quad \chi^-_1\to \chi^0_1 e^-\bar\nu_e
\quad{\rm``deep-inelastic\ sneutrino-chargino"}
\end{array}\right.\nonumber
\end{eqnarray}
All the above processes are kinematically accessible in the $SU(5)\times U(1)$
models. In the minimal $SU(5)$ model the sleptons are heavy and neither the
slepton pair production at LEPII nor any of the indicated processes at HERA
are allowed.

We now give a sample of the actual results obtained for the most important
processes listed above. These are shown in
Figures~\ref{Figure6},\ref{Figure7a},\ref{Figure7},\ref{Figure8}.
%\vspace{1cm}
\begin{figure}[p]
\vspace{4.5in}
\includegraphics{fig6Layman.ps}
\vspace{2.6in}
\includegraphics{fig7Layman.ps}
\vspace{-2.2in}
\caption{\baselineskip=12pt
The number of trilepton events at the Tevatron per $100\ipb$ in the minimal
$SU(5)$ model and the no-scale $SU(5)\times U(1)$ model (for $m_t=130\GeV$).
Note that with $200\ipb$ and 60\% detection efficiency it should be possible to
probe basically all of the parameter space of the minimal $SU(5)$ model, and
probe chargino masses as high as $175\GeV$ in the no-scale model. Upper bounds
on the trilepton cross section ($\sigma\cdot B<(0.6-1)$ pb) have been recently
announced by the CDF [40] and D0 [41] Collaborations.}
\label{Figure6}
\end{figure}

\begin{figure}[p]
\vspace{4.5in}
\includegraphics{fig8Layman.ps}
\vspace{-2.5in}
\caption{\baselineskip=12pt
The number of ``mixed" events (1-lepton+2jets+$\mpt$) events per ${\cal
L}=100\ipb$ at LEPII versus the chargino mass in the minimal $SU(5)$ model.}
\label{Figure7a}
\vspace{5in}
\includegraphics{fig9Layman.ps}
\vspace{-0.3in}
\caption{\baselineskip=12pt
The number of ``mixed" events (1-lepton+2jets+$\mpt$) events per ${\cal
L}=100\ipb$ at LEPII versus the chargino mass in the no-scale model (top row).
Also shown (bottom row) are the number of di-electron events per ${\cal
L}=100\ipb$  from selectron pair production versus the lightest selectron
mass.}
\label{Figure7}
\end{figure}

\begin{figure}[t]
\vspace{4.7in}
\includegraphics{fig10Layman.ps}
\vspace{-0.3in}
\caption{\baselineskip=12pt
The elastic plus deep-inelastic total supersymmetric cross section at HERA
($ep\to{\rm susy}\to eX+\mpt$) in the no-scale model versus the lightest
selectron mass ($m_{\tilde e_R}$) and the sneutrino mass ($m_{\tilde\nu}$). The
short- and long-term limits of sensitivity are expected to be $10^{-2}\pb$ and
$10^{-3}\pb$ respectively.}
\label{Figure8}
\end{figure}

More recently it has been realized that indirect experimental constraints
on the $SU(5)\times U(1)$ models exist and can be quite significant. These
constraints come generally from three sources:
(i) the experimentally allowed range for  $B(b\to s\gamma)$ as recently
determined by the CLEO Collaboration \cite{bsgamma+bsg-eps}; (ii) the
long-standing value for the anomalous magnetic moment of the muon \cite{g-2};
and (iii) the precision LEP measurements of the electroweak parameters in the
form of the allowed range for the $\epsilon_1$ parameter \cite{ewcorr}. (Only
the last constraint is relevant in the minimal $SU(5)$ supergravity model
because of its relatively heavier spectrum.)

The first two constraints exclude regions of the parameter space of the
$SU(5)\times U(1)$ models which span all allowed values of the chargino
mass, when viewing the parameter space in the $(m_{\chi^\pm_1},\tan\beta)$
plane for fixed $m_t$. On the other hand, the $\epsilon_1$ constraint
basically implies an upper bound on $m_t$: $m_t\lsim165\GeV$, unless the
chargino is very light ($m_{\chi^\pm_1}\lsim70\GeV$) in which case the
upper bound on $m_t$ can be relaxed up to $m_t\lsim180\GeV$. As an example
of the effect of the constraints, in Fig.~\ref{Figure9} we show the parameter
space for $m_t=130\GeV$ (when the $\epsilon_1$ constraint is not restrictive)
for the no-scale model \cite{All}. Clearly the region of parameter space
accessible to LEPII searches ($m_{\chi^\pm_1}\lsim100\GeV$) has become quite
constrained.

\vspace{1.5cm}
\begin{figure}[t]
\vspace{4.7in}
\includegraphics{fig11Layman.ps}
\vspace{-1.5in}
\caption{\baselineskip=12pt
The parameter space of the no-scale $SU(5)\times U(1)$ supergravity model
for $m_t=130\GeV$. Points denoted by periods satisfy all presently known
experimental constraints, whereas those denoted by pluses violate the
limits on $B(b\to s\gamma)$ and those denoted by crosses violate the limits
on $(g-2)^{susy}_\mu$.}
\label{Figure9}
\end{figure}

\vfill\eject

\section{Conclusions}
In sum, we have presented a simplified tour of supersymmetric unified theories
which hopefully will allow non-experts in the field to get acquainted with
such a topical subject. One of our goals was to show that supersymmetric
particles can well be ``around the corner" and at the verge of discovery
at present and near future facilities, such as the Tevatron, HERA, and LEP (I
and II). We have also shown that there are many experimental constraints on
supersymmetric unified models which need to be consistently imposed to speak
about experimentally viable models. Finally, these experimentally testable
models should be as simple as the best theoretical motivations allow. There
is little to be learned from generic models whose many parameters can be tuned
to predict anything. In fact, predicting anything is akin to predicting
nothing.

\section*{Acknowledgements}
This work has been supported in part by DOE grant DE-FG05-91-ER-40633.
%\vfill\eject

\end{document}